\PassOptionsToPackage{square}{natbib}
\PassOptionsToPackage{sort&compress}{natbib}
\PassOptionsToPackage{numbers}{natbib}

\documentclass[final,5p,times,twocolumn]{elsarticle}
 \makeatletter
    \def\ps@pprintTitle{%
      \let\@oddhead\@empty
      \let\@evenhead\@empty
      \let\@oddfoot\@empty
      \let\@evenfoot\@oddfoot
    }
    \makeatother
\usepackage{graphicx}
\usepackage{float}
\usepackage{amssymb}
\usepackage{upgreek}
\usepackage[square,sort&compress,numbers]{natbib}
 \usepackage{lineno}

  \def\km{\kern-1.5mm}

\newcommand{\affuni}[2]{Dipartimento di Fisica dell'Universit\`a #1, #2, Italy.}
\newcommand{\affinfn}[2]{INFN Sezione di #1, #2, Italy.}

\journal{Phisics Letter B}

\begin{document}

\begin{frontmatter}



%\title{Combined limit on the production of a new vector boson in continuum processes at KLOE}
\title{Combined limit on the production of a light gauge boson decaying into $\mu^+\mu^-$ and $\pi^+\pi^-$}

\author{The KLOE-2 Collaboration}
\author[Messina,Frascati]{\\A.~Anastasi}
\author[Frascati]{D.~Babusci}
\author[Frascati,Warsaw]{M.~Berlowski}
\author[Frascati]{C.~Bloise}
\author[Frascati]{F.~Bossi}
\author[INFNRoma3]{P.~Branchini}
\author[Roma3,INFNRoma3]{A.~Budano}
\author[Uppsala]{B.~Cao}
\author[Roma3,INFNRoma3]{F.~Ceradini}
\author[Frascati]{P.~Ciambrone}
\author[Frascati]{F.~Curciarello\corref{mycorrespondingauthor}}
\cortext[mycorrespondingauthor]{Corresponding author}
\ead{francesca.curciarello@lnf.infn.it}
\author[Cracow]{E.~Czerwi\'nski}
\author[Roma1,INFNRoma1]{G.~D'Agostini}
\author[Frascati]{E.~Dan\`e}
\author[INFNRoma2]{V.~De~Leo}
\author[Frascati]{E.~De~Lucia}
\author[Frascati]{A.~De~Santis}
\author[Frascati]{P.~De~Simone}
\author[Roma3,INFNRoma3]{A.~Di~Cicco}
\author[Roma1,INFNRoma1]{A.~Di~Domenico}
\author[Frascati]{D.~Domenici}
\author[Frascati]{A.~D'Uffizi}
\author[Roma2,INFNRoma2]{A.~Fantini}
\author[GSSI]{G.~Fantini}
\author[Frascati]{P.~Fermani}
\author[ENEAFrascati,INFNRoma1]{S.~Fiore}
\author[Cracow]{A.~Gajos}
\author[Roma1,INFNRoma1]{P.~Gauzzi}
\author[Frascati]{S.~Giovannella}
\author[INFNRoma3]{E.~Graziani}
\author[BINP,Novosibirsk]{V.~L.~Ivanov}
\author[Uppsala]{T.~Johansson}
\author[Frascati]{X.~Kang}
\author[Cracow]{D.~Kisielewska-Kami\'nska}
\author[BINP,Novosibirsk]{E.~A.~Kozyrev}
\author[Warsaw]{W.~Krzemien}
\author[Uppsala]{A.~Kupsc}
\author[BINP,Novosibirsk]{P.~A.~Lukin}
\author[Messina2,INFNCatania]{G.~Mandaglio \corref{mycorrespondingauthor}}
\ead{gmandaglio@unime.it}
\author[Frascati,Marconi]{M.~Martini}
\author[Roma2,INFNRoma2]{R.~Messi}
\author[Frascati]{S.~Miscetti}
\author[INFNRoma2]{D.~Moricciani}
\author[Cracow]{P.~Moskal}
\author[INFNRoma3]{A.~Passeri}
\author[Energetica,INFNRoma1]{V.~Patera}
\author[Frascati]{E.~Perez~del~Rio}
\author[INFNRoma2]{N.~Raha}
\author[Frascati]{P.~Santangelo}
\author[Calabria,INFNCalabria]{M.~Schioppa}
\author[Roma3,INFNRoma3]{A.~Selce}
\author[Cracow]{M.~Silarski}
\author[Frascati,IFIN]{F.~Sirghi}
\author[BINP,Novosibirsk]{E.~P.~Solodov}
\author[INFNRoma3]{L.~Tortora}
\author[INFNPisa]{G.~Venanzoni}
\author[Warsaw]{W.~Wi\'slicki}
\author[Uppsala]{M.~Wolke}
\address[INFNCatania]{\affinfn{Catania}{Catania}}
\address[Cracow]{Institute of Physics, Jagiellonian University, Cracow, Poland.}
\address[Frascati]{Laboratori Nazionali di Frascati dell'INFN, Frascati, Italy.}
\address[GSSI]{Gran Sasso Science Institute, L'Aquila, Italy.}
\address[IFIN]{Horia Hulubei National Institute of Physics and Nuclear Engineering, M\v{a}gurele, Romania}
\address[Messina]{Dipartimento di Scienze Matematiche e Informatiche, Scienze Fisiche e Scienze della Terra dell'Universit\`a di Messina, Messina, Italy.}
\address[Messina2]{Dipartimento di Scienze Chimiche, Biologiche, Farmaceutiche ed Ambientali dell'Universit\`a di Messina, Messina, Italy.}
\address[BINP]{Budker Institute of Nuclear Physics, Novosibirsk, Russia.}
\address[Novosibirsk]{Novosibirsk State University, Novosibirsk, Russia.}
\address[INFNPisa]{\affinfn{Pisa}{Pisa}}
\address[Calabria]{\affuni{della Calabria}{Rende}}
\address[INFNCalabria]{INFN Gruppo collegato di Cosenza, Rende, Italy.}
\address[Energetica]{Dipartimento di Scienze di Base ed Applicate per l'Ingegneria dell'Universit\`a ``Sapienza'', Roma, Italy.}
\address[Marconi]{Dipartimento di Scienze e Tecnologie applicate, Universit\`a ``Guglielmo Marconi", Roma, Italy.}
\address[Roma1]{\affuni{``Sapienza''}{Roma}}
\address[INFNRoma1]{\affinfn{Roma}{Roma}}
\address[Roma2]{\affuni{``Tor Vergata''}{Roma}}
\address[INFNRoma2]{\affinfn{Roma Tor Vergata}{Roma}}
\address[Roma3]{Dipartimento di Matematica e Fisica dell'Universit\`a 
``Roma Tre'', Roma, Italy.}
\address[INFNRoma3]{\affinfn{Roma Tre}{Roma}}
\address[ENEAFrascati]{ENEA, Department of Fusion and Technology for Nuclear Safety and Security, Frascati (RM), Italy}
\address[Uppsala]{Department of Physics and Astronomy, Uppsala University, Uppsala, Sweden.}
\address[Warsaw]{National Centre for Nuclear Research, Warsaw, Poland.}

\begin{abstract}
 We searched for the $\mu^+\mu^-$ decay of a light vector gauge boson, also known as dark photon, in the $e^+ e^- \to \mu^+ \mu^- \gamma_{\rm ISR}$ process by means of the Initial State Radiation (ISR) method.  We used  1.93~fb$^{-1}$ of data collected by the KLOE experiment at the DA$\Phi$NE $\phi$-factory.
No structures have been observed over the irreducible $\mu^+ \mu^-$ background. A 90\% CL limit on the ratio $\varepsilon^2=\alpha^{\prime}/\alpha$ between the dark coupling constant and the fine structure constant of $ 3\times 10^{-6}-2\times 10^{-7}$ has been set in the dark photon mass region between 519 MeV and 973 MeV. This new limit has been combined with the published 
result obtained investigating the hypothesis of the dark photon decaying into hadrons in $e^+ e^- \to  \pi^+  \pi^- \gamma_{\rm ISR}$ 
events. 
The combined 90\% CL limit increases the sensitivity especially in the $\rho-\omega$ interference region and excludes $\varepsilon^2$ greater than $(13-2)\times 10^{-7}$. 
For dark photon masses greater than 600 MeV the combined limit is lower than 8~$\times\, 10^{-7}$ resulting more stringent than present constraints from other experiments.

\end{abstract}


\end{frontmatter}
%\clearpage
 %\linenumbers
%\tableofcontents
%\clearpage
\section{Introduction}
\label{Introduction}

Many gravitational anomalies observed since the first decades of the twentieth century, as well as large-scale structure formation in the early Universe, can be explained by 
the existence of a non-baryonic matter known as dark matter (DM)~\cite{PDG}.
Dark matter motivates extending the Standard Model of particle physics (SM) to include a dark sector consisting
of  fields and particles  with  no  SM  gauge  charges and including  extra gauge symmetries.  
The minimal extension of the SM consists of just one additional abelian gauge symmetry $U_{\rm D}(1)$ with associated a light vector gauge boson, the dark photon -- known also as $U$ boson, $\gamma^{\prime}$ or $A^{\prime}$--  as mediator of the new force, called for this reason dark force.
In the simplest scenario~\cite{Holdom}, the coupling with SM particles arises from a vector portal known as kinetic mixing  consisting in loops of heavy dark particles charged under both the electromagnetic and the dark force.
The portal allows the mixing of the dark photon belonging to the $U_{\rm D}(1)$ group with the SM photon of the  $U_{\rm em}(1)$ symmetry introducing the Lagrangian term:
\begin{equation}
L_{mix}=-\frac{\varepsilon}{2} \, F_{ij}^{\rm em} \, F^{ij}_{ \rm dark}.
\end{equation}
Here  $\varepsilon$ is a dimensionless parameter which governs the strength of the mixing 
($\varepsilon^2 = \alpha^{\prime}/\alpha$, $\alpha=\alpha_{em}, \, \alpha^{\prime}$ is the effective dark coupling constant) while $F_{ij}^{\rm em}$ and $F^{ij}_{\rm dark}$ are  the field strength tensors of the SM $U_{\rm em}(1)$ and dark $U_{\rm D}(1)$ gauge groups, respectively. Through the portal the $U$ boson can couple to the electromagnetic current with a strength proportional to the SM particles electric charge. The process is responsible for both production and decay of the dark photon in SM interactions thus resulting in an $\varepsilon^2$ suppression.
If the kinetic mixing appears at
the one-loop level, $\varepsilon$ can be estimated to be in the range $10^{-2}-10^{-6}$ allowing visible effects at high luminosity  $e^+ e^-$ colliders~\cite{Essig}.

During the last decade, the dark photon has been the focus of a world-wide intensive research because considered as possible explanation of many astrophysical puzzling evidences~\cite{Pamela}.\\
In this work we investigate the simplest hypothesis of a visibly decaying dark photon looking for resonant production of $U$ boson from the continuum, considering as allowed only decays into SM particles. 
 The $U$ signal should appear as a peak in the invariant mass  of the final state particles with a width mainly dominated by the  invariant mass resolution since the expected $U$-decay width can be considered negligible~\cite{babayaga_article}. 
  KLOE already investigated 
 $e^+ e^- \to U h^\prime$ (dark Higgsstrahlung)~\cite{KLOE_Hi}, $U$ boson in decays of vector particles to pseudoscalars~\cite{KLOE_UL1,KLOE_UL2}, and
 the visible decay hypothesis publishing three searches for radiative $U$ production in the $e^+ e^- \to U \gamma$ process, with the $U$ boson decaying into: a) $\mu^+ \mu^-$ ~\cite{mmg}, using 240~pb$^{-1}$ of data; b) $e^+ e^-$~\cite{eeg}, using a sample of 1.54~fb$^{-1}$; c) $\pi^+ \pi^-$ ~\cite{ppg} analyzing the whole KLOE data set corresponding to   an integrated luminosity of 1.93~fb$^{-1}$.  Searches for muon and pion pairs, with the ISR photon selected at small angle   ($\theta~<~15^{\circ},\, \theta~>~165^{\circ}$), cover approximately the same $U$-boson mass range of 520--990 MeV, while for the electron pairs the photon selection was at large angle ($55^{\circ}~<~\theta~<~125^{\circ}$)  allowing to reach a lowest $U$-boson mass of 5 MeV and probing the $(g-2)_{\mu}$ favoured region~\cite{a_mu}. \\ In the present work we extend the statistics of the $U \to \mu^+ \mu^-$ search to the whole data sample and update the analysis with a new estimate of the background,  analogous to the one used for  the  $U \to \pi^+\pi^-$  search. The new search confirms no $U$-boson signal  in the dimuon invariant mass spectrum: a new 90\% CL exclusion limit  in $\varepsilon^2$ is estimated. This limit is of comparable magnitude with respect to the previous ones,  thus a combined search of dark photon decays into both muon and pion pairs 
 would increase the sensitivity of the single channel searches, particularly, it is more effective in the region of the $\rho-\omega$ interference where the search for $U \to \mu^+ \mu^-$ loses sensitivity. 
\section{ The KLOE detector}
\label{sec:kloe_detector}

The KLOE detector operates at DA$\Phi$NE\cite{daf}, the Frascati $\phi$-factory. DA$\Phi$NE is an $e^+ e^-$  collider  working at a center of mass energy $ m_\upphi\simeq 1.019$ GeV. Positron and electron beams collide at an angle of $\pi-$25 mrad, producing $\phi$ mesons nearly at rest. The detector consists of a large cylindrical drift chamber (DC)~\cite{KLOE_DC}, surrounded by a lead scintillating-fiber electromagnetic calorimeter (EMC)~\cite{KLOE_EMC}.  
A superconducting coil around  the EMC provides a 0.52 T magnetic field along the bisector of the colliding beams which is taken as the $z$ axis of our coordinate
system. 

The EMC barrel and end-caps cover 98\% of the solid angle. Calorimeter modules are read out at both ends by 4880 photomultipliers. Energy and time resolutions are $ \sigma_E /E=~0.057 /\sqrt{E(\rm{GeV})} $ and $ \sigma_t =57\ \rm{ps}/\sqrt{E(\rm{GeV})}\oplus 100\ \rm{ps}$, respectively. 
The drift chamber has only stereo wires and is $4$ m in diameter, $ 3.3$ m long. It is built out of carbon-fibers and operates with a low-$Z$ gas mixture (helium with 10\% isobutane). Spatial resolutions are  $\sigma_{xy}\sim 150\ \rm\upmu m$ and  $\sigma_z\sim 2$ mm. The momentum resolution for large angle tracks is $\sigma(p_\perp) / p_\perp\sim 0.4\% $. The trigger uses both EMC and DC information. Events used in this analysis are triggered by at least two energy deposits larger than 50~MeV in two sectors of the barrel calorimeter~\cite{KLOE_trig}.

\section{$e^+ e^- \to  \mu^+  \mu^- \gamma$ data analysis}
\label{Data Analysis}

\subsection{Event Selection}
\label{selection}
We selected $\mu^+\mu^-\gamma$ candidates by requiring events with two oppositely-charged tracks emitted at large polar angles, $50^{\circ}~<~\theta~<~130^{\circ}$, with the  undetected ISR photon missing momentum pointing -- according to the $\mu^+\mu^-\gamma$ kinematics --  at small polar angles ($\theta~<~15^{\circ},\, \theta~>~165^{\circ}$).
The tracks are required to have the point of closest approach to the $z$  axis within a cylinder of radius 8 cm and length 15 cm centered at the interaction point. 
In order to ensure good reconstruction and efficiency, we selected tracks with transverse and longitudinal momentum ${p}_\perp  >$~160~MeV or $|p_{z}| > $~90~MeV, respectively. This separation of track and photon selection regions in the analysis, greatly reduces the contamination from the resonant process $e^+e^-\to \phi\to\pi^+\pi^-\pi^0$, from the Final State Radiation (FSR) processes $e^+e^-\to \pi^+\pi^-\gamma_{\rm FSR}$ and $e^+e^-\to \mu^+\mu^-\gamma_{\rm FSR}$, since the $\mu^+\mu^-\gamma$ cross section diverges at small ISR photon angle making FSR processes and $\phi$ decays relatively unimportant~\cite{Binner,Kloe05,KLOE1,Kloe12}.
Consequently, since ISR-photons are mostly collinear with the beam line, a high   statistics for the ISR signal events remains.
The main background contributions affecting the ISR $\mu^+\mu^-\gamma$  sample are the resonant $ e^+ e^- \to\phi \to  \pi^+  \pi^-  \pi^0$ process and the ISR and FSR $e^+e^- \to  x^+ x^- \upgamma(\upgamma),\, x=\mathrm{e},\pi$ processes.
\begin{figure}[htp!]
\begin{center}\vspace{-0.2cm}
\includegraphics[width=8.5cm]{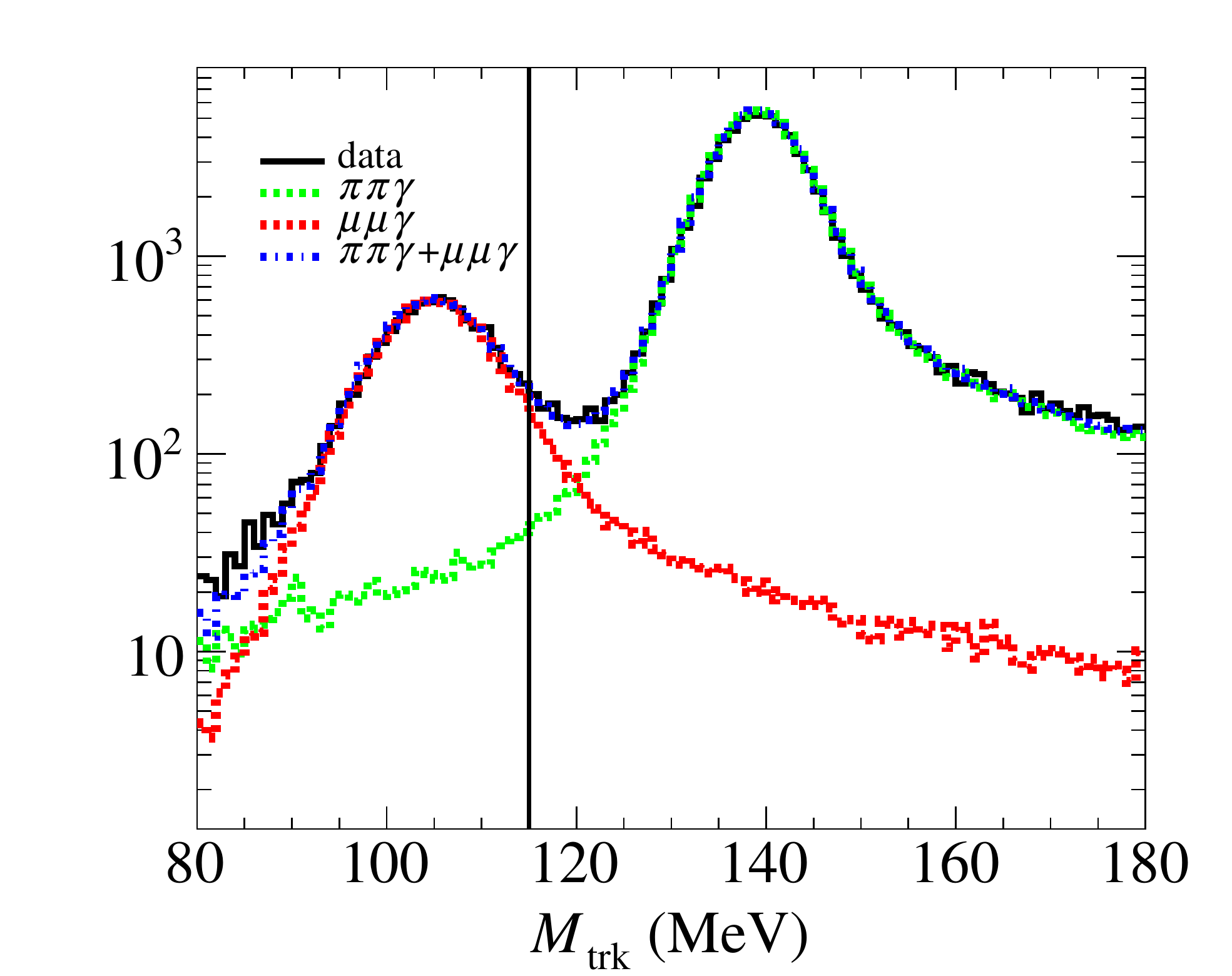}
\caption{$M_{\rm trk}$ distributions for $ \mu^+  \mu^+ \gamma$ and $ \pi^+  \pi^- \gamma$. Data are represented in black, the MC simulations of $ \pi^+ \pi^-\gamma$ and  $ \mu^+ \mu^-\gamma$ channels are in green and red, respectively, while their sum is in blue; the vertical back line represents the selection cut applied to separate the two channels.}
\label{mtrkcut}
\end{center}
\end{figure}
Their contributions have been evaluated by applying kinematical cuts in the $M_{\rm trk}$, $M^2_{\pi\pi}$ plane\footnote{$M_{\rm trk}$ is computed from energy
and momentum  conservation, assuming the presence of one undetected photon
and that the tracks belong to particles of the same mass: 
$$
\left(\sqrt{s}-\sqrt{|\vec{p}_+|^2 + M^2_{\rm trk}}-
\sqrt{|\vec{p}_-|^2 + M^2_{\rm trk} }\right)^2-\left(\vec{p}_+
+\vec{p}_-\right)^2 = 0
$$
where $\vec{p}_+$ ($\vec{p}_-$) is the measured momentum of the positive (negative) particle, and only one of the four solutions is physical.}, with $M_{\pi\pi}$ the invariant mass of the track pair in the pion
mass hypothesis.

\begin{figure}[htp!]
\begin{center}
\includegraphics[width=8.5cm,height=6.8cm]{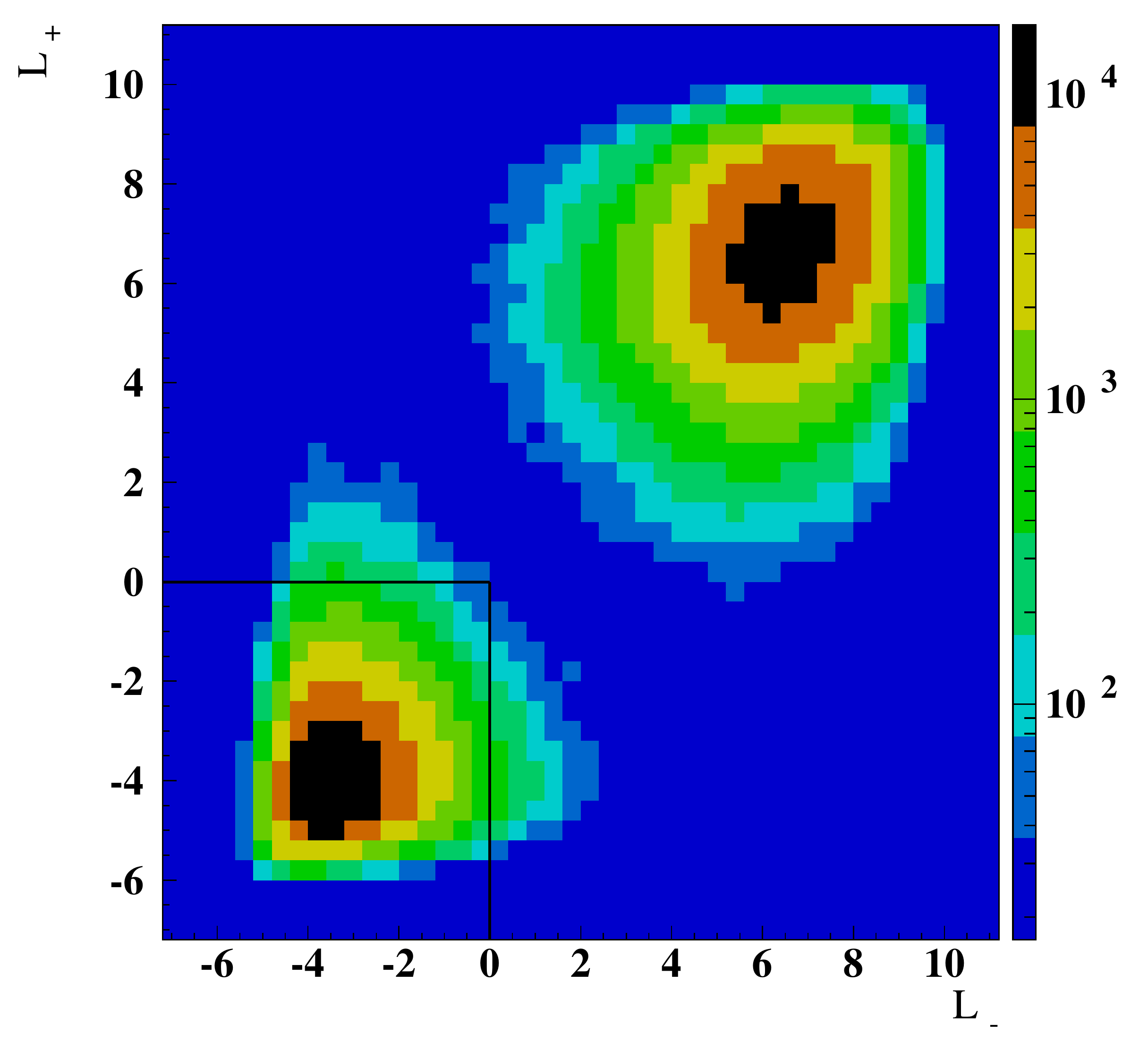}
\caption{MC $L_{+}$ vs. $L_{-}$ PID distributions
for both tracks. Events contained in the low left rectangle (having  both tracks with $L_{\pm}<0$) are regarded as $e^+e^-\gamma$ events
and rejected in the selection.}
\label{lpm}
\end{center}
\end{figure}

A particle identification estimator (PID), $L_{\pm}$, based on a pseudo-likelihood function  using the charged particles time-of-flight and energy depositions in the five calorimeter layers is used to suppress radiative Bhabha events~\cite{memo,KLOE1,KLOE2}.  Events with both tracks having $L_{\pm} < 0$ are identified as $e^+  e^- \gamma$ events and rejected (see Figure \ref{lpm}). 

Finally, a cut on the track-mass variable $M_{\rm trk}$ selects muons by requiring $M_{\rm trk}~<~ 115$~MeV as shown in Figure~\ref{mtrkcut}.
At the end of the selection described above about 7.16~$\times\, 10^{6}$ events survive. 

\begin{figure}[ht!]
\begin{center}
\includegraphics[width=8.5 cm]{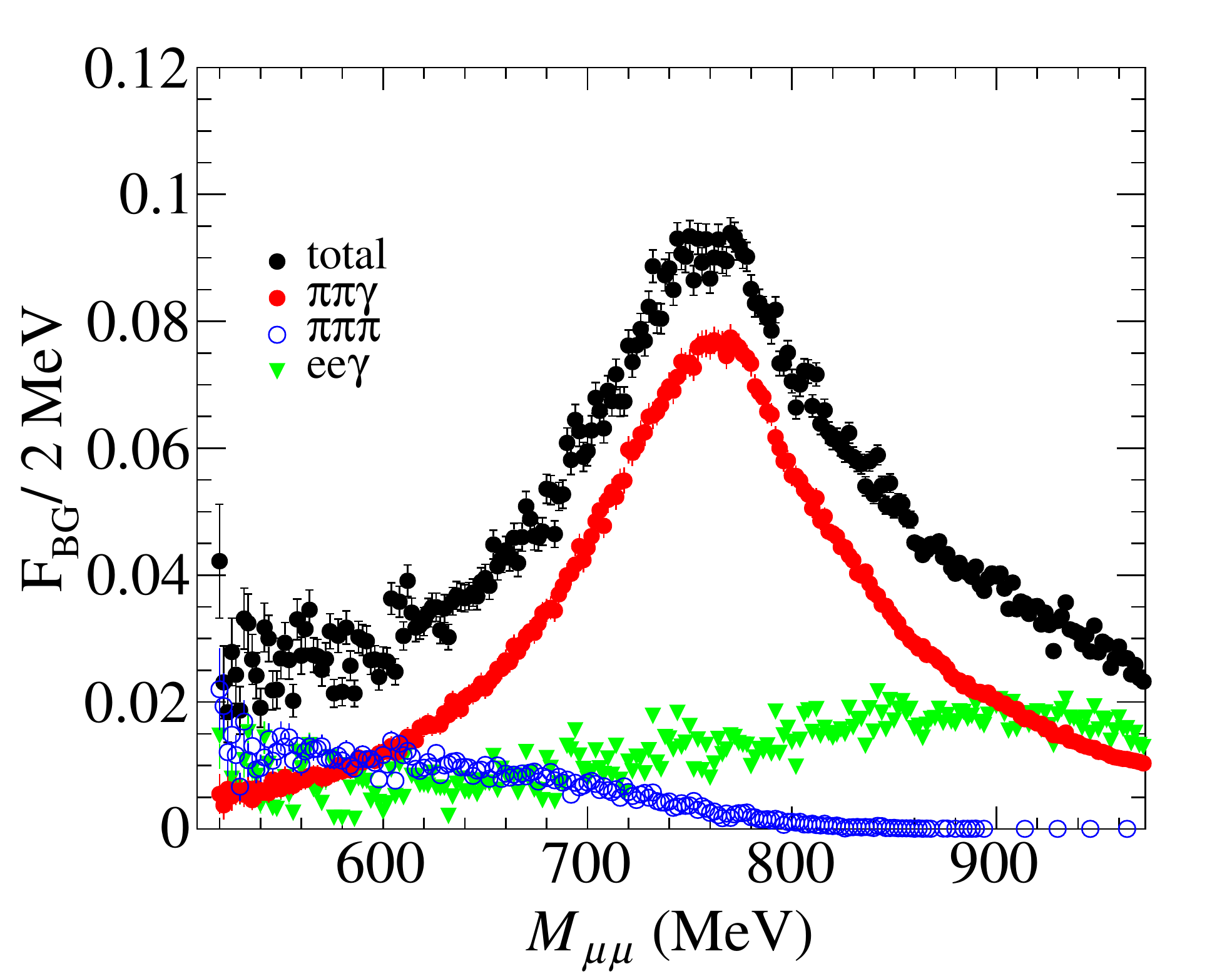}
\caption{Fractional residual backgrounds as function of $M_{ \mu \mu}$.}
\label{Bckg}
\end{center}
\end{figure}

In order to evaluate the residual background contributions, the same analysis chain was applied  to simulated events for the $\pi^+  \pi^- \gamma$ and $\pi^+  \pi^-  \pi^0$ channels while the radiative Bhabha contribution has been evaluated directly from measured data. Distributions of the fractional residual background $F_{\rm BG}$ for each channel and their sum  
are shown in Figure ~\ref{Bckg} as a function of the invariant mass of the track pair in the muon mass hypothesis, $M_{\mu\mu}$.

The total residual background rises up to about 9\%  in the $\rho-\omega$ region and decreases down to about 3\% at low and high invariant mass values.

\section{Parametrization of the irreducible $\mu^+\mu^-\gamma$ background}
\label{parback}

\begin{figure}[ht!]
\begin{center}
\includegraphics[width=8. cm]{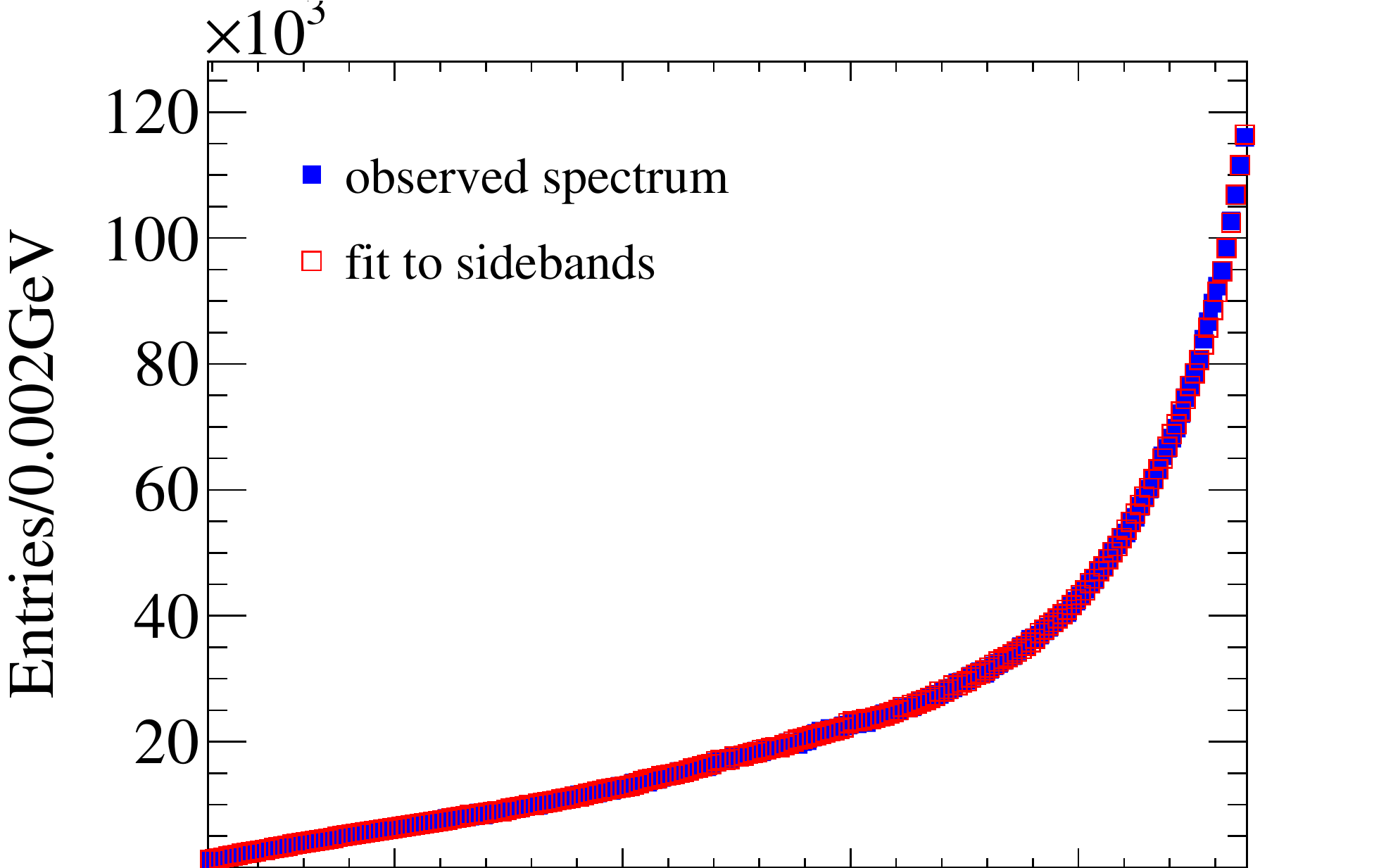} \\
\includegraphics[width=8. cm]{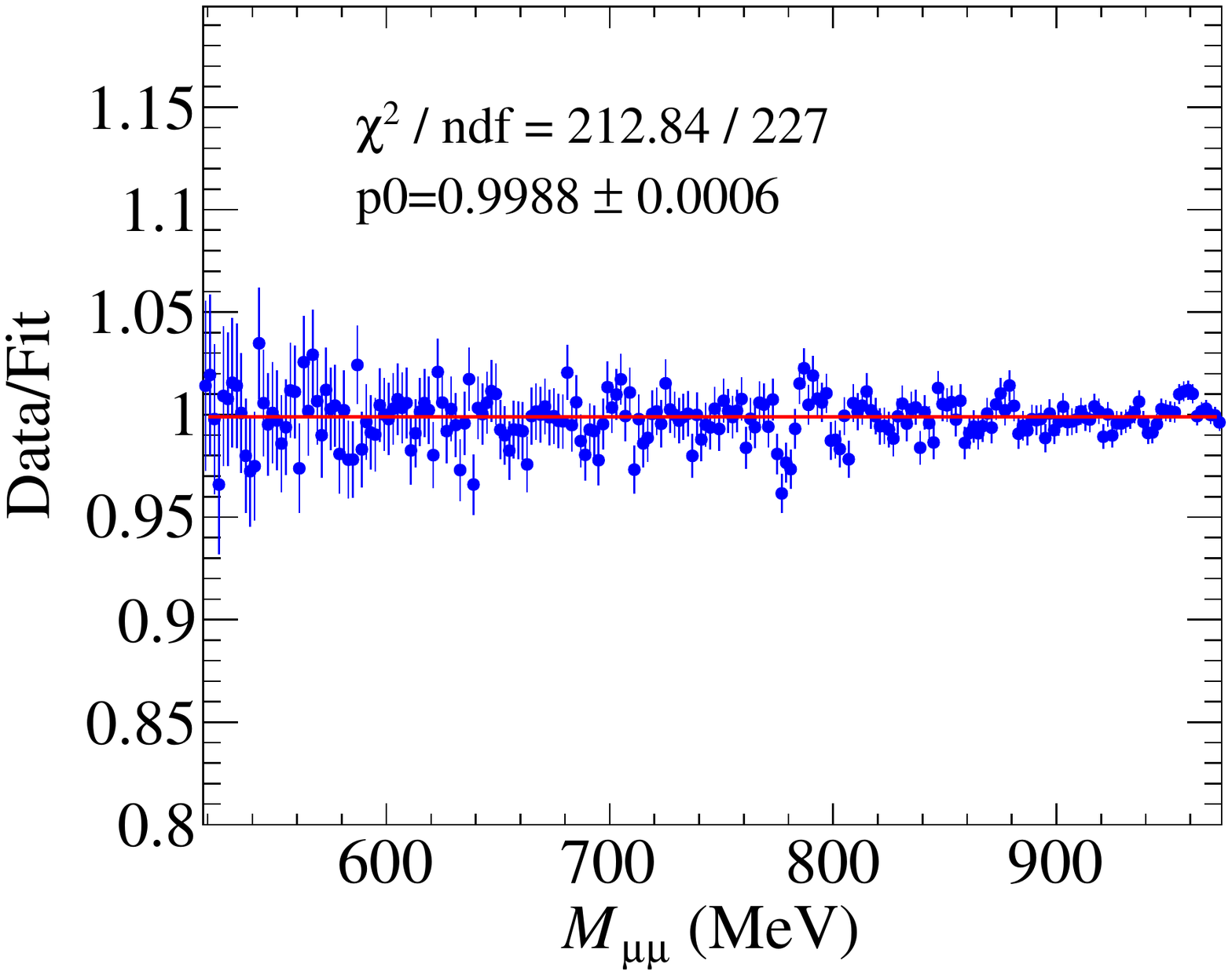}
\caption{Top panel:  $\mu^+\mu^-\gamma$  observed spectrum (full squares) and estimated irreducible
background (open squares). Bottom panel: data and estimated background
ratio.}
\label{spectrum}
\end{center}
\end{figure}

To minimize the systematic uncertainties affecting the analysis, we evaluated the irreducible $\mu^+ \mu^- \gamma$ background directly from the data. In Figure~\ref{spectrum}, we report the comparison between  data and  estimated background distributions (top panel) and their ratio (bottom panel), which are in good agreement within errors.

\begin{figure}[ht!] 
\centering \vspace{-0.2cm}
\includegraphics[width=7.8cm]{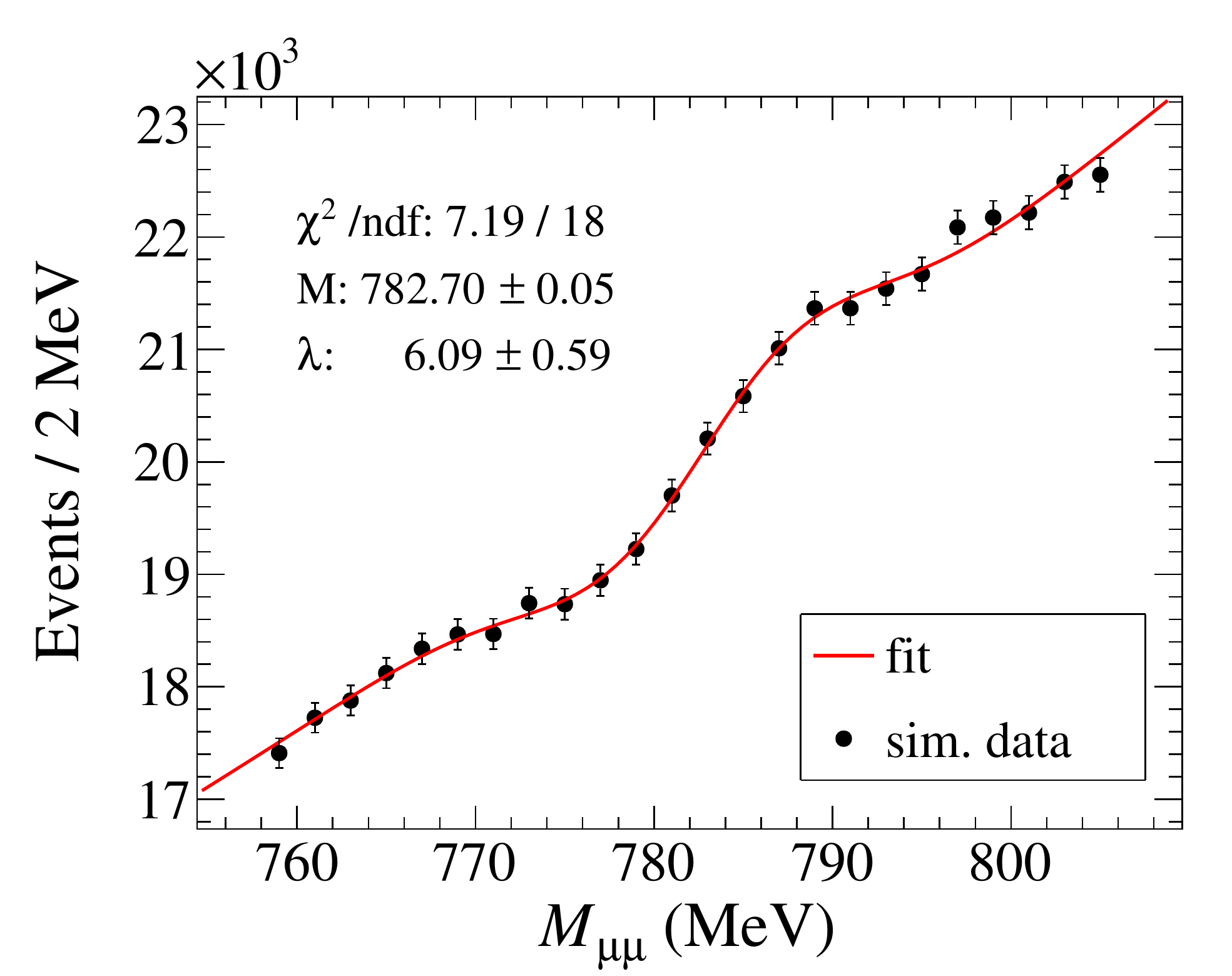}
\caption{Fit of reconstructed $\mathtt{PHOKHARA}$ MC with vacuum polarisation correction included.\label{validation}}
\end{figure}

We estimated the irreducible $\mu^+ \mu^- \gamma$  background by using a side band fit to the observed spectrum, keeping, for each iteration, the fit with the best reduced $\chi^2$.
The fit to side bands in the whole mass range has been performed considering sub
ranges $\pm12\sigma$ wide, where $\sigma$ is the dimuon invariant mass resolution of about 2 MeV \cite{ppg}. For each U-mass hypothesis a region corresponding to $\pm3\sigma$  is excluded from the fit.
We fit the data distributions by using Chebyshev polynomials (as in Ref. [9]) up to 6th order in the mass ranges 519--757 MeV and 811--973 MeV. In the mass interval between 759 and 809 MeV, where the effect of the $\rho-\omega$ interference is present \cite{KLOE_VP}, we used another parametrization:
\begin{center}
\begin{equation}
{\rm f(x) = pol2(x)}\,\cdot\, [1 + A\cdot(x-M)\cdot\exp(-0.5\cdot{((x-M)/\lambda)}^2)].\,\,
\label{parVp}
\end{equation}
\end{center}

The parametrization (\ref{parVp}) has been used because found to best fit
the $\mu\mu$ invariant mass simulated spectrum (PHOKHARA generator ~\cite{H,H_1,H_2,PHOKHARA} with vacuum polarisation correction included and a full description of the detector performed with the GEANFI package~\cite{GEANFI}) as shown in Figure \ref{validation}. 
As a first step, the three coefficients of the second order polynomial pol2(x) and the parameters $A$, $M$ and $\lambda$ are computed by fitting the function in Eq. \ref{parVp} over the full $\mu^+ \mu^- \gamma$  simulated spectrum: values of 782.24 MeV and 6.09 MeV were obtained for the parameters $M$ and $\lambda$, respectively. Then, the fits in the considered mass range (759--809 MeV) of the $\mu^+\mu^-\gamma$ observed spectrum have been performed by using again the function (\ref{parVp}),  keeping the parameters $M$ and $\lambda$ fixed at the values 782.24 MeV and 6.09 MeV, and leaving free all the other parameters.

\begin{figure}[ht!]
\begin{center}
\hspace{0.4cm}\includegraphics[width=7.8cm]{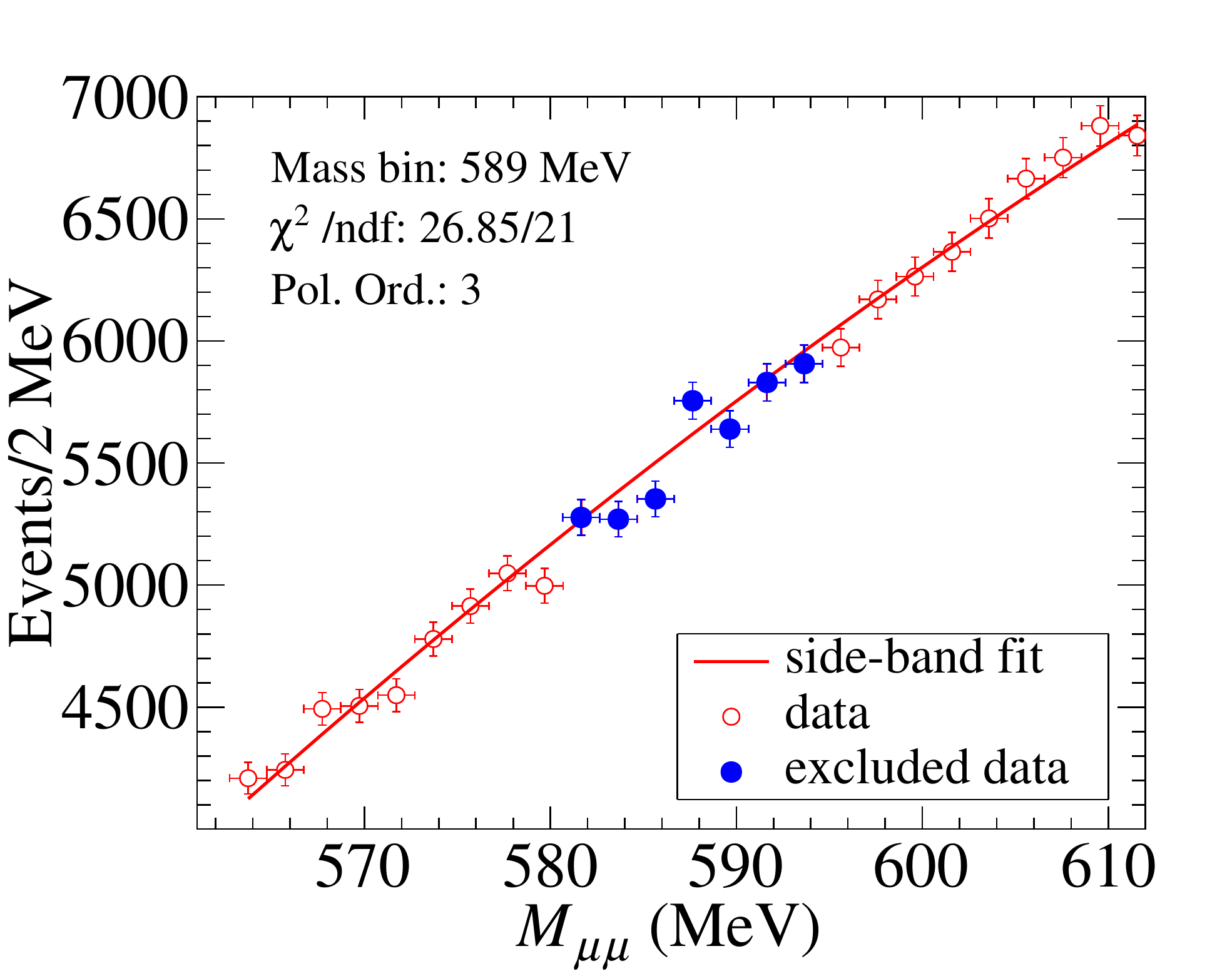}\\
\includegraphics[width=7.8cm]{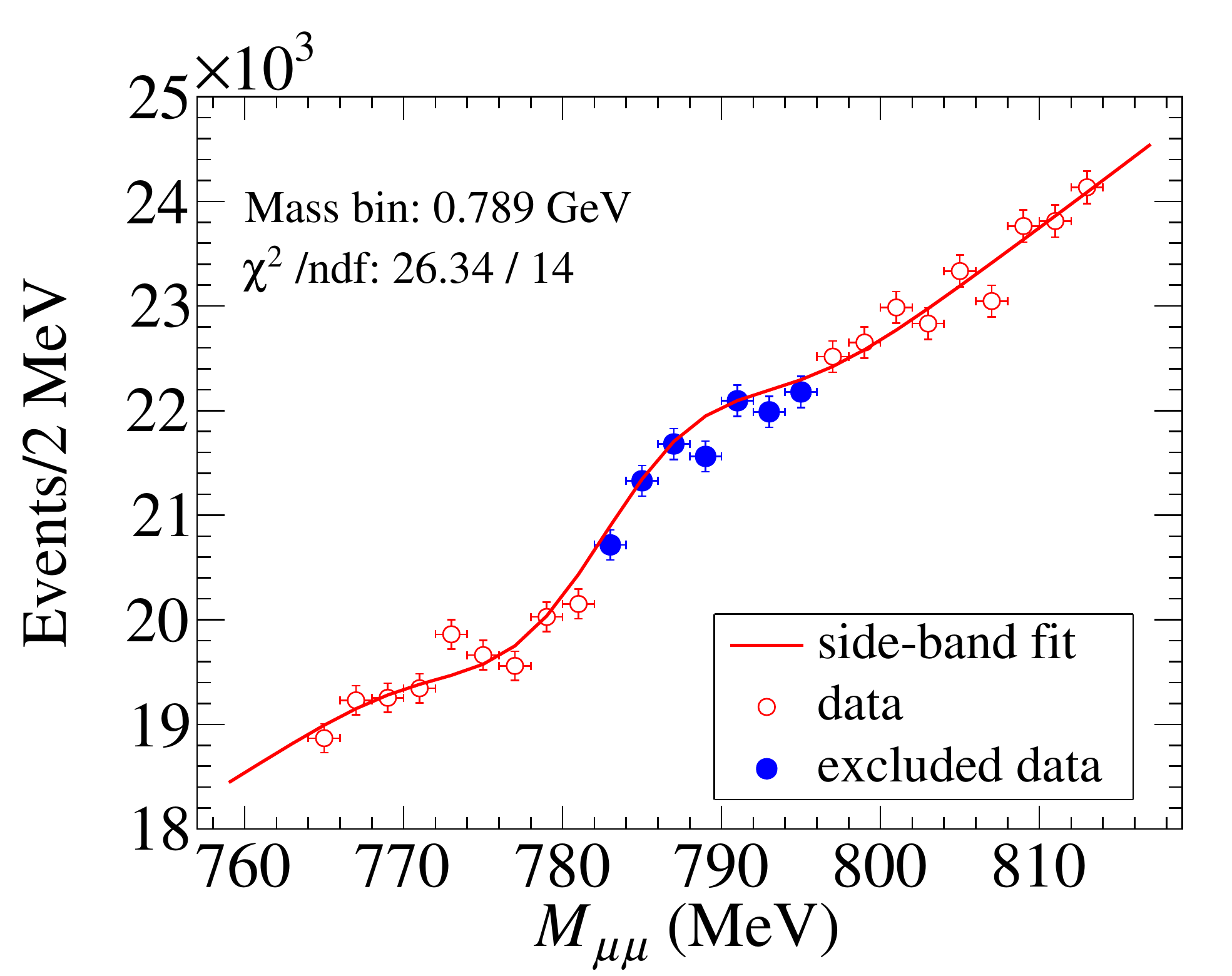}
\caption{Examples of fits performed in two sub-ranges of the $ \mu^+  \mu^- \gamma$  spectrum by using Chebyshev polynomials (upper panel) and parametrization (\ref{parVp})  (lower panel).}
\label{fit_firstrange}\vspace{-0.4cm}
\end{center}
\end{figure}

Examples of the fits performed by using Chebyshev polynomials or the parametrization in eq. (\ref{parVp}) are shown in Figure \ref{fit_firstrange}.

The reduced $\chi^2$ of the fit to side bands for both parameterizations remains below 2 in the whole mass range. The fit procedure is stable in the whole data range and no anomaly is observed in the fitted background.

\section{Systematic uncertainties}
\label{section_syst}
In the following we report the systematic uncertainties affecting the analysis,
mainly due to the evaluation of the irreducible background and to the event
selection applied to the $\mu^+ \mu^- \gamma$  candidates. 
\subsection{Systematic uncertainties on the irreducible background}
\begin{figure}[htb!]
\begin{center}
\includegraphics[width=8.2cm]{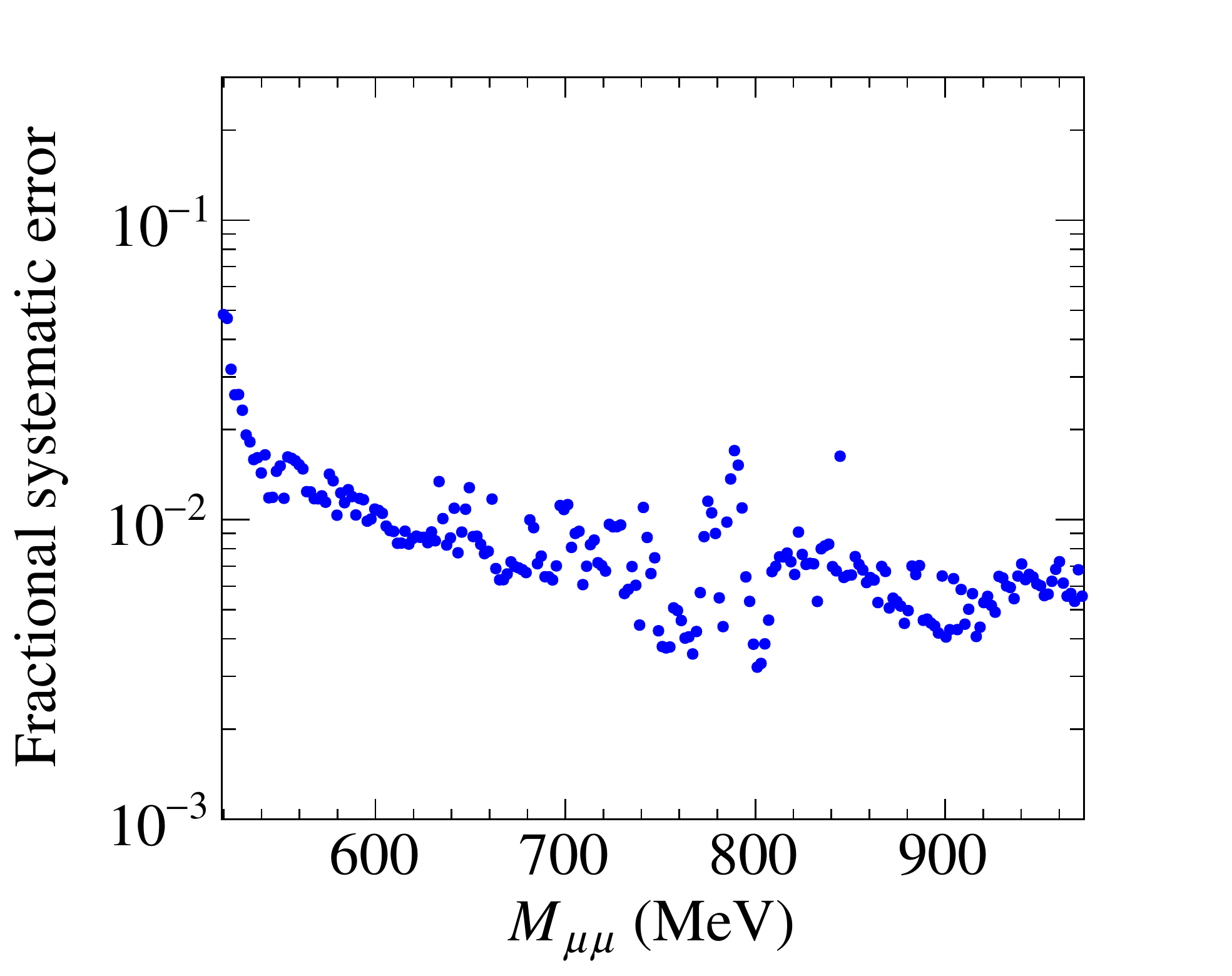}
\caption{Bin-by-bin total fractional systematic error of the background estimate.}
\label{tot_tlimit_syst}
\end{center}
\end{figure}

The fractional systematic error on the irreducible $\mu^+\mu^-\gamma$ background is shown in Figure~\ref{tot_tlimit_syst}. 
The evaluation of the systematic uncertainties has been derived for each mass bin by estimating the error of the fit.
The total systematic error is less than 1\% in most of the mass range.

The systematic error due to the side bands fit procedure has been also evaluated by varying the range of the fit interval of $\pm1\sigma$ and computing the maximum difference between nominal fit and the fit derived by changing the fit interval. Its contribution is $<<1$\% and therefore results negligible in the whole mass range.  

\subsection{Systematic uncertainties of the global efficiency}
\begin{figure}[htp!]
\begin{center}
\includegraphics[width=8.2 cm]{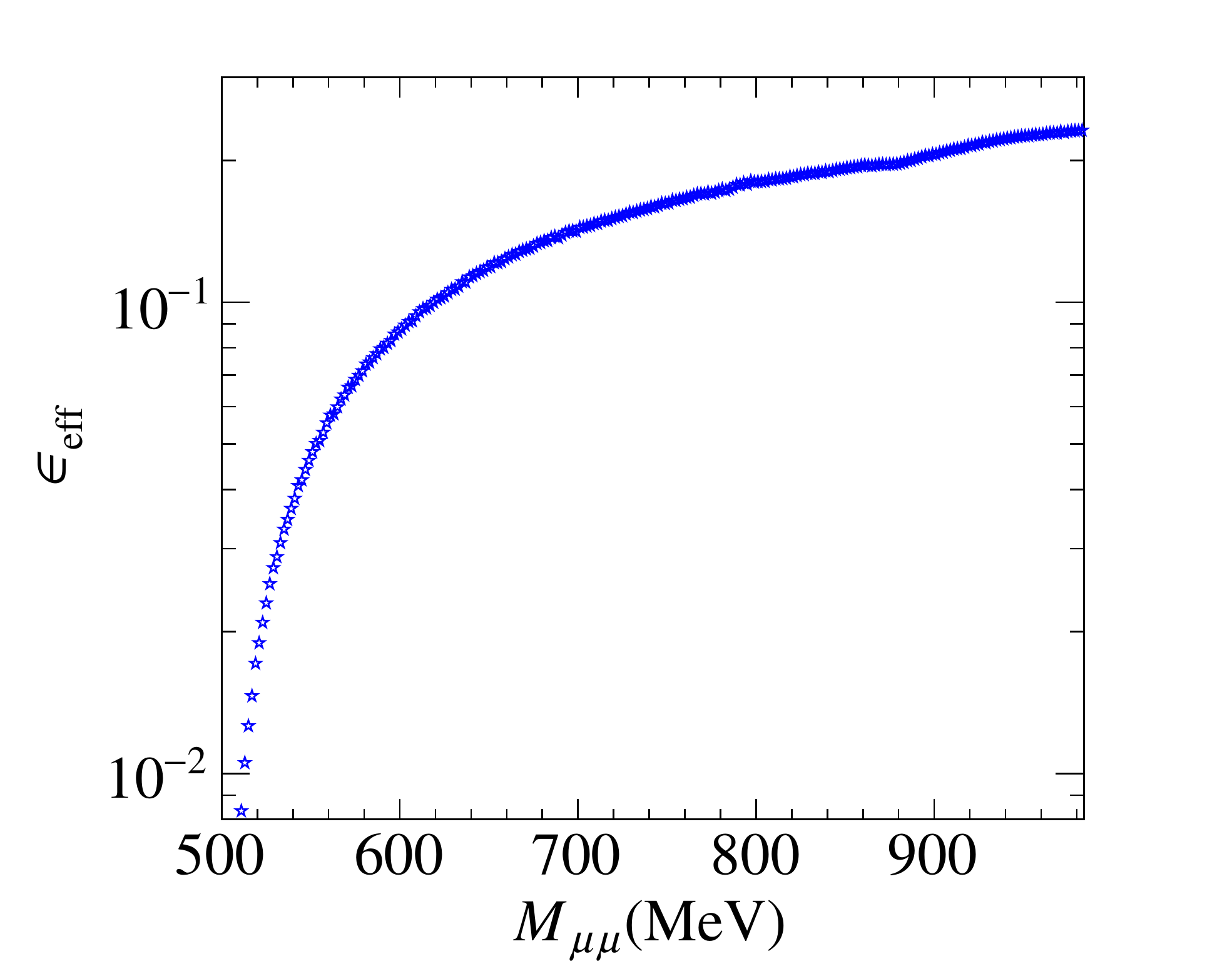}
\caption{Global efficiency as function of $M_{\mu\mu}$.}
\label{global_eff}
\end{center}
\end{figure}

Figure~\ref{global_eff} shows the global analysis efficiency that has been evaluated from a full  $ \mu^+  \mu^- \gamma$ simulation. 
This efficiency includes contributions from kinematic selection, trigger, tracking, acceptance and PID-likelihood efficiencies.

Table~\ref{tab:1} lists all the systematic errors affecting the $\mu^+\mu^-\gamma$ analysis.
We evaluated the 
corresponding uncertainties by using the same procedures described in Ref.~\cite{mmg}. These systematic uncertainties do not affect the irreducible background estimation but enter in the determination of the selection efficiency and the luminosity measurement.
\begin{table}[ht]
\begin{center}
\caption{Summary of the systematic uncertainties.} 
\label{tab:1}\begin{small}
\begin{tabular}{cc} 
\hline
Systematic source  & $\qquad$ Relative uncertainty (\%)   \\
\hline
 $M_{\rm trk}$ cut                    & $\qquad$ 0.4 \\
Acceptance               & $\qquad$ 0.6 -- 0.1 as $M_{\mu\mu}$ increases \\
%\hline                                   
Trigger                  & $\qquad$ 0.1   \\
%\hline 
Tracking                 &  $\qquad$ 0.3 -- 0.6 as $M_{\mu\mu}$ increases   \\
%\hline
Generator                 & $\qquad$ 0.5 \\
%\hline
Luminosity               & $\qquad$ 0.3    \\
%\hline
 PID                     & $\qquad$ negligible    \\
 %\hline
Total                 &  $\qquad$ 0.98 -- 0.94 as $M_{\mu\mu}$ increases\\
 \hline
\end{tabular}
\end{small}
\end{center}
\end{table}

\section{Limits on $U$-boson production in $\mu\mu\gamma$ events}

 The  $\mu^+\mu^-\gamma$ observed spectrum does not reveal the presence of any  visible structure (see Figure~\ref{spectrum}) within the mass-dependent systematic uncertainties.  
 For this reason, a procedure has been applied to evaluate the statistical significance of the observed data fluctuations and eventually set a limit on  the ${e^+ e^-}\to U \gamma,\, U \to \mu^+\mu^-$ process.  
The following subsection describes the results of the limit extraction procedure.

\subsection{Upper Limit Extraction on $\varepsilon^2$}
\label{UL}

To extract the upper limit (UL) on $\varepsilon^2$ we used the Confidence Level Signal (CL$_{\rm S}$) technique~\cite{CLS_Technique}.
The procedure requires as inputs the invariant mass data spectrum, the background (the irreducible $\mu^+ \mu^- \gamma$ background), the $U$-boson signal and the systematic  fractional uncertainties on the background estimation for each $M_{\mu\mu}$ bin.
The signal has been generated with a toy MC in steps of 2 MeV for the $U$-boson mass. At each step, a Gaussian distribution is built with a width corresponding to the invariant mass resolution of the dimuon system of about 2~MeV.
The signal is then integrated over $M_{\mu\mu}$ around $M_U$.
The number of signal events, given as input to the procedure, is initially arbitrary and very high (about ten times the square root of the estimated background value in the corresponding mass bin) and then iteratively scaled until the confidence level $\rm CL_S$ reaches 0.1 within $\pm$~0.01.  
The integral of the signal corresponding to the defined level of confidence represents the limit on the number of $U$-boson events excluded at 90\% CL.
 Since the limit is strongly dependent on the irreducible background evaluation, the limit extraction accounts for the systematic uncertainties of the background estimate. 
 The limit extraction procedure uses the total bin-by-bin fractional systematic uncertainty, reported in Figure~\ref{tot_tlimit_syst}, to perform a Gaussian smearing of the $\mu^+\mu^-\gamma$ expected background given as input.

The UL on the kinetic mixing parameter has been extracted by using, for each $U$-boson mass value,  the following formula~\cite{mmg,eeg,ppg}:
\begin{equation}
\varepsilon^2=\frac{\alpha^{\prime}}{\alpha}= \frac{N_{ \rm CLS}}{\epsilon_{\rm eff} \cdot L \cdot H \cdot I}
\label{eq.3}
\end{equation}
where $N_{\rm CLS}$ is the limit on the number of % $U$-boson 
events, $\epsilon_{\rm eff}$ represents the global efficiency (shown in Figure~\ref{global_eff}), $L$ is the integrated luminosity (1.93~fb$^{-1}$ with an uncertainty of 0.3\%\cite{Kloe05,KLOE1}), $H$ is the radiator function calculated at QED next-to-leading-order corrections with an uncertainty of 0.5\%~\cite{H,H_1,H_2,H_3} and given by:
\begin{equation}
H=\frac{\mathrm{d}\sigma_{ \mu  \mu \gamma}/\mathrm{d} M_{\mu\mu}}{\sigma(e^+ e^-\rightarrow  \mu^+  \mu^-,M_{\mu\mu} )}\,.
\label{eq.4}
\end{equation}
Here $\mathrm{d}\sigma_{ \mu  \mu \gamma}/\mathrm{d}M_{\mu\mu}$ is the differential cross section of $e^+e^-\rightarrow \mu^+ \mu^-\gamma$, $\sigma(e^+ e^-\rightarrow  \mu^+  \mu^-, M_{\mu\mu})$ is the total cross section of the $e^+e^-\rightarrow \mu^+ \mu^-$ process. In Eq. (\ref{eq.3}), $I$ is given by the following integral around $M_U$:
\begin{equation}
I=\int \sigma^{ \mu  \mu}_U \, \mathrm{d}\sqrt{s}\,,
\label{eq.5}
\end{equation}
where $\sigma^{ \mu  \mu}_U = \sigma(e^+ e^- \rightarrow U\rightarrow  \mu^+  \mu^-, s)$ is the total cross section of $U$-boson production decaying in the $ \mu^+  \mu^-$ channel  when the  kinetic mixing parameter $\varepsilon$ is equal to 1, $s=M_U^2$.
The uncertainties on $H$, 
$\epsilon_{\rm eff}$, $L$, and $I$, propagate to the systematic error on $\varepsilon^2$ via eq.(\ref{eq.3}). The resulting uncertainty on $\varepsilon^2$ is lower than 1\% and has been taken into account in the estimated limit.

\begin{figure}[htp!]
\begin{center}
\includegraphics[width=9cm]{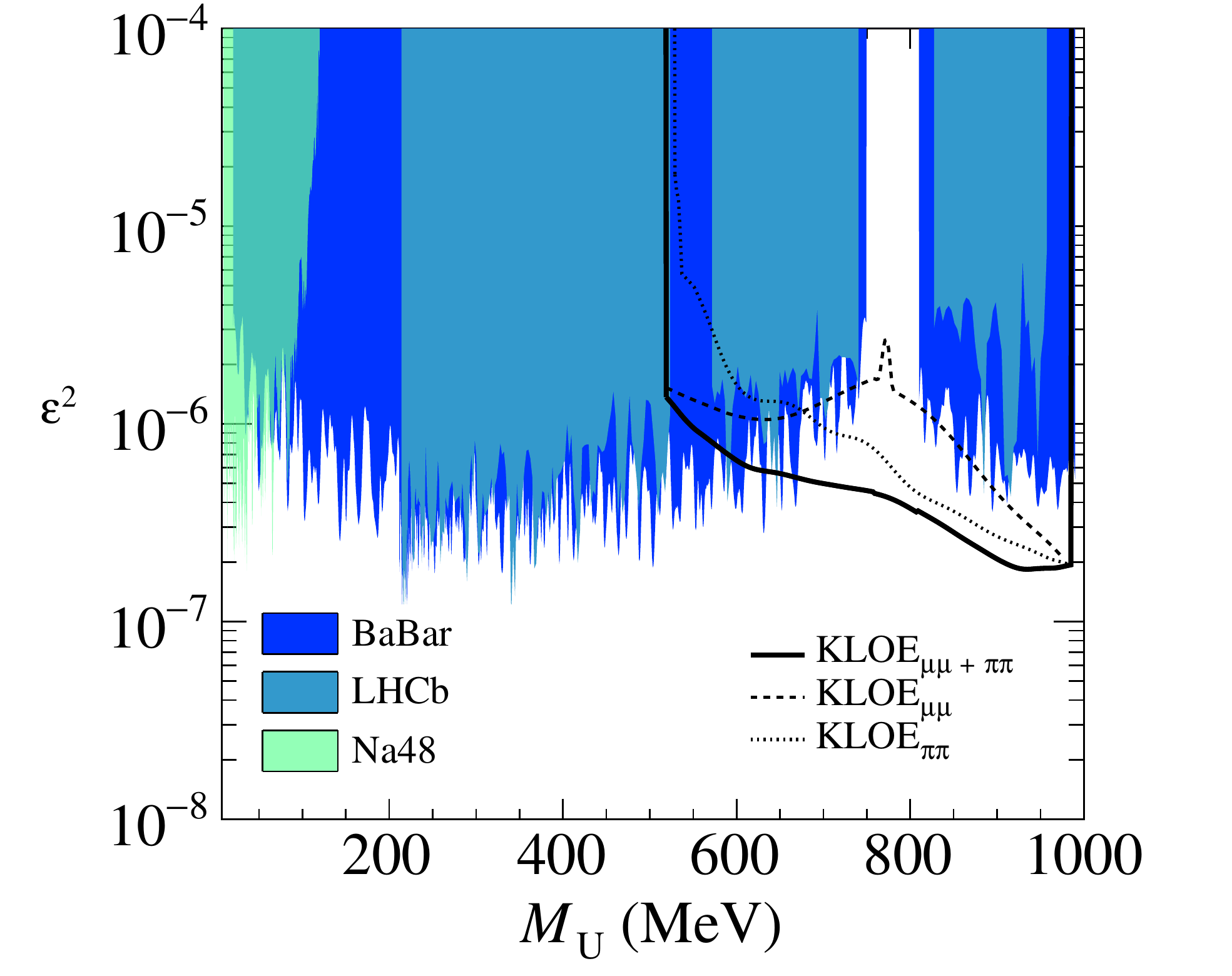}\vspace{-0.2cm}
\caption{90\% CL exclusion plot  for $\varepsilon^2$ as a function of the $U$-boson mass for the ${e^+ e^-}\to U \gamma$\, process. The  $U \to\mu^+\mu^-$ limit (dashed line), the $U\to\pi^+\pi^-$ ~\cite{ppg} constraint (dotted line), and the $U\to \mu^+\mu^-,\pi^+\pi^-$ combination (solid line) at full KLOE statistics, are presented in comparison with the competitive limits by BaBar~\cite{BaBar},  NA48/2~\cite{NA48/2}, and LHCb experiments \cite{lhcb}.}\label{Exclusion_plot_comby}\vspace{-0.5cm}
\end{center}
\end{figure}

The exclusion plot on $\varepsilon^2$ is shown as a dashed line in Figure~\ref{Exclusion_plot_comby} compared with the existing limits in the mass range below $1$~GeV. Our 90\% CL UL ranges from $ 3\times 10^{-6}$  to $2 \times 10^{-7}$ in the 519--973 MeV mass interval.

\section{Combined limit on $U$-boson production in $\mu\mu\gamma$ and $\pi\pi\gamma$ events}

In this section we present the combination procedure of the full statistics $\pi^+\pi^-\gamma$ and $\mu^+\mu^-\gamma$ limits. As for the previous analyses, we use the CL$_{\rm S}$ technique to estimate a 90\% CL limit for the ${e^+e^-}\to U \gamma_{\rm ISR},\, U \to \mu^+\mu^-,\pi^+\pi^-$ process.  To extract the limit, we use the already estimated  background and observed spectra for both $\pi\pi\gamma$ \cite{ppg} and  $\mu\mu\gamma$  channels in a combined way. 
A total systematic error on the irreducible background estimate,  given by the combination of the corresponding estimated uncertainties for both $U$-boson decay modes, is also given as input to the procedure.
A combined $U$-boson signal is generated for both decay channels taking into account  the differences in global efficiency and relative branching ratio~\cite{Essig}. The signal inputs are generated with the same toy MC procedure performed for the $\mu^+\mu^-\gamma$ limit extraction, then, each signal is integrated and normalised to the number of events estimated from Eq. (\ref{eq.3}), for a given hypothesis of the kinetic mixing parameter $\varepsilon^2$. 
The limit computation proceeds according to the following steps: it makes a hypothesis of the $\varepsilon^2$ kinetic mixing parameter, starting from an arbitrary very low value; the corresponding number of events for $\pi\pi\gamma$ and $\mu\mu\gamma$ channels are generated according to Eq. (\ref{eq.3}) in order to build the signal input histogram,  then, the procedure runs as before by comparing data and expected irreducible background. The search procedure ends when the estimated CL$_{\rm S}$ becomes close to 0.1 within $\pm$0.01, providing directly the corresponding exclusion on  $\varepsilon^2$. 

The combined upper limit, obtained after averaging the statistical fluctuations by a smoothing procedure,  excludes values of  $\varepsilon^2$ greater than  $(13-2) \times 10^{-7}$ in the $U$-mass range 519--987 MeV. It is  shown in Figure~\ref{Exclusion_plot_comby}, compared to the most competitive limits. The other existing 
limits \cite{Mami1,Apex,KLOE_UL1,KLOE_UL2,mmg,eeg,WASA,HADES} are not reported to make the figure more readable. The combined limit is represented by the blue area and is more stringent with respect to the already set limits in the mass region 600--987~MeV, while it is comparable to BaBar and LHCb results for masses lower than 600~MeV.

\section{Conclusions}
\label{conclusions}

We analyzed  1.93 fb$^{-1}$ of KLOE data to investigate the hypothesis of a light vector gauge boson decaying into muons and pions by means of the ISR method in the  $e^+e^-\rightarrow U \gamma_{\rm ISR},\, U \to \mu^+\mu^-,\pi^+ \pi^-$ process. No $U$-boson evidence has been found and a combined limit at 90\% CL using the two  $U$-decay modes has been extracted on the  kinetic mixing parameter $\varepsilon^2$ in the energy range between 519 and 987~MeV. The new combined limit is more stringent than the already set constraints in the region between 600 and 987  MeV by excluding values of  $\varepsilon^2$ higher than $(8-2)\times10^{-7}$.   

\section*{Acknowledgments}
\label{sec:acknowledgments}
We warmly thank our former KLOE colleagues for the access to the data collected during the KLOE data taking campaign.
We thank the DA$\Phi$NE team for their efforts in maintaining low background running conditions and their collaboration during all data taking. We want to thank our technical staff: 
G.F. Fortugno and F. Sborzacchi for their dedication in ensuring efficient operation of the KLOE computing facilities; 
M. Anelli for his continuous attention to the gas system and detector safety; 
A. Balla, M. Gatta, G. Corradi and G. Papalino for electronics maintenance;  
C. Piscitelli for his help during major maintenance periods. 
This work was supported in part  
by the Polish National Science Centre through the Grants No.\
2013/11/B/ST2/04245,
2014/14/E/ST2/00262,\\
2014/12/S/ST2/00459,
2016/21/N/ST2/01727,\\
2016/23/N/ST2/01293,
2017/26/M/ST2/00697.

\end{document}